\documentstyle[twocolumn,aps]{revtex}
\begin{document}
\draft
\preprint{}
\title{Quantum Nondemolition Monitoring of Universal Quantum Computers}
\author{Masanao Ozawa}
\address{School of Informatics and Sciences,
Nagoya University, Nagoya 464-01, Japan}
\date{\today}
\maketitle
\begin{abstract}
The halt scheme for quantum Turing machines, originally proposed by 
Deutsch, is reformulated precisely and is proved to work 
without spoiling the computation. 
The ``conflict'' pointed out recently by Myers in the definition of 
a universal quantum computer is shown to be only apparent.
In the context of quantum nondemolition (QND) measurement,
it is also shown that the output observable, an observable representing 
the output of the computation, is a QND observable and that 
the halt scheme is equivalent to the QND 
monitoring of the output observable.
\end{abstract}
\pacs{PACS numbers:  89.70.+c, 03.65.-w, 03.65.Bz}
\narrowtext

Since Shor \cite{Sho94} found efficient quantum algorithms
for the factoring problem and the discrete logarithm problem,
which are considered to have no efficient algorithms in computational 
complexity theory and applied to public-key cryptosystems \cite{Sti95},
a great deal of attention has focused on quantum computation
\cite{EJ96}.
The notion of quantum algorithm is currently defined through two different
approaches: (1) quantum Turing machines \cite{Deu85,BV93}
and (2) quantum circuits \cite{Deu89,Yao93}.
While a circuit represents a single algorithm for a set of 
size-limited input data, a universal Turing machine \cite{Tur36} 
models a programmable computing machine which can 
compute every recursive function by reading the program as a part 
of the input data.  In quantum computation, Deutsch \cite{Deu85}, 
Bernstein and Vazirani \cite{BV93} showed constructions of universal 
quantum Turing machines.
In a recent article Myers \cite{Mye97} pointed out, however, 
a ``conflict'' in the definition of a quantum Turing machine and 
questioned the computational power of universal quantum Turing machines.
The purpose of this letter is to show that the ``conflict'' is only
apparent.

A Turing machine is a deterministic discrete-time dynamical 
system including a bilateral infinite tape and a head to read 
and write a symbol on the tape.  
Its configuration is determined by the internal state $q$ of the machine,
the position $h$ of the head on the tape, 
and the symbol string $T$ on the tape.
For any integer $i$ the symbol in the cell at the position $i$ on 
the tape is denoted by $T(i)$.
If $C$ denotes a configuration of a Turing machine,
the internal state, the head position, and the tape string 
in the configuration $C$ are denoted by $q_{C}$, $h_{C}$, and $T_{C}$,
respectively; and hence we can write 
$C=(q_{C},h_{C},T_{C})$.
The time evolution of the Turing machine is determined 
by the (classical) {\em transition function} $\delta_{c}$
such that the relation
$\delta_{c}(p,\sigma)=(\tau,q,d)$
represents the transition of the configutation that if the internal state is
$p$ and if the head reads the symbol $\sigma$ then
the head writes the symbol $\tau$ on the tape,
the internal state turns to $q$, and the head moves one cell to
the direction $d=+1$ (right) or $d=-1$ (left).
Thus, if $\delta_{c}(q_{C},T_{C}(h_{C}))=(\tau,q,d)$, 
the configuration $C=(q_{C},h_{C},T_{C})$ changes as
\begin{equation}
(q_{C},h_{C},T_{C}) \to (q,h_{C}+d,T^{\tau}_{C}),
\end{equation}
where $T^{\tau}_{C}$ denotes the tape string such that
$T^{\tau}_{C}(i)=T_{C}(i)$
if $i\ne h_{C}$, and that $T^{\tau}_{C}(h_{C})=\tau$.

A quantum Turing machine is the quantization of a classical 
Turing machine.
Its quantum {\em state} is represented by a vector in the Hilbert space 
spanned by the computational basis, a complete orthonormal 
system in one-to-one correspondence with the set of configurations 
of the classical machine.
Thus, the computational basis is represented by 
\begin{equation}
|C\rangle=|q_{C}\rangle|h_{C}\rangle|T_{C}\rangle
\end{equation}
for any configuration $C$ of the classical machine.
The time evolution of the quantum Turing machine
is described by a unitary operator $U$ determined
by the complex-valued {\em quantum transition function}
such that the relation
$\delta(p,\sigma,\tau,q,d)=c$ represents that the classical transition 
$(p,\sigma)\mapsto(\tau,q,d)$ occurs with the amplitude $c$.
Thus the time evolution operator $U$ is determined by
\begin{eqnarray}
\lefteqn{U|q_{C}\rangle|h_{C}\rangle|T_{C}\rangle}\quad\nonumber\\
&=&\sum_{\tau,q,d}\delta(q_{C},T_{C}(h_{C}),\tau,q,d)
|q\rangle|h_{C}+d\rangle|T^{\tau}_{C}\rangle.
\end{eqnarray}

The result of a computation is obtained by measuring the tape string
after the computation has been completed.
Unlike the classical case, the machine configuration cannot be 
monitored throughout the computation because of the inevitable
disturbance caused by measurement.
Thus, the machine needs a specific halt scheme
to signal actively when the computation has been completed.
The halt scheme proposed by Deutsch \cite{Deu85} is as follows.
Deutsch introduced an additional single qubit, called the 
{\em halt qubit}, together with an observable $\hat{n}_{0}$, called 
the {\em halt flag}, with the eigenstates $|0\rangle$ and $|1\rangle$, 
so that the internal state $q$ is represented by the state vector 
$|q\rangle|1\rangle$ if $q$ is the final state in the classical picture or by 
$|q\rangle|0\rangle$ otherwise.
Then, the halt qubit is initialized to $|0\rangle$ before starting the 
computation, 
and every valid quantum algorithm sets the halt qubit to $|1\rangle$ 
when the computation has been completed
but does not interact with the halt qubit otherwise.
Deutsch claimed that 
{\em the observable $\hat{n}_{0}$ can then be periodically observed 
from the outside without affecting the operation of the machine}.

Recently, Myers \cite{Mye97} argued that 
a measurement of the halt qubit might spoil the computation. 
Myers's argument, with some modifications, runs as follows.
After $N$ steps of computation
the machine is, 
in general, in such a superposition of states of the computational
basis as
\begin{equation}\label{eq:324b}
\sum_{\alpha}c_{\alpha}
|q_{\alpha}\rangle|h_{\alpha}\rangle|T_{\alpha}\rangle|0\rangle
+
\sum_{\beta}c_{\beta}
|q_{\beta}\rangle|h_{\beta}\rangle|T_{\beta}\rangle|1\rangle.
\end{equation}
If after each step we have either $c_{\alpha}=0$ for all $\alpha$
or $c_{\beta}=0$ for all $\beta$, then the halt flag can be measured
repeatedly during a computation without changing the state
and {\em a forteori} without spoiling the computation.
But, because of quantum parallelism \cite{Deu85,Jos91},
there should be cases where we have some $N_{B}\gg N_{A}$ such that
after $N$ steps with $N_{A}<N<N_{B}$ neither of the above holds.
Then the state (\ref{eq:324b}) can be written as
\begin{mathletters}
\begin{equation}
c_{A}|A\rangle|0\rangle+c_{B}|B\rangle|1\rangle,
\end{equation}
where 
\begin{eqnarray}
c_{A}|A\rangle
&=&\sum_{\alpha}c_{\alpha}
|q_{\alpha}\rangle|h_{\alpha}\rangle|T_{\alpha}\rangle\not=0,\\
c_{B}|B\rangle
&=&\sum_{\beta}c_{\beta}
|q_{\beta}\rangle|h_{\beta}\rangle|T_{\beta}\rangle\not=0.
\end{eqnarray}
\end{mathletters}
In this range of steps the state entangles the non-halt
qubits with the halt qubits, so that the measurement of
the halt flag changes the state and, Myers concluded, 
spoils the computation.

In what follows, it will be proved that the 
measurement of the halt flag, though changes the state,
does {\em not} spoil the computation so that
the halt scheme works even in the entangled case.   

The precise formulation of the halt scheme is given as follows.

(I) The halt flag $\hat{n}_{0}$ is measured instantaneously
after every step.  
This measurement is a precise measurement
of the observable $\hat{n}_{0}$ satisfying the {\em projection postulate},
i.e., the measurement changes the state as follows:
\begin{eqnarray*}
\lefteqn{c_{A}|A\rangle|0\rangle+c_{B}|B\rangle|1\rangle}\quad\nonumber\\
&\to&
\left\{
\begin{array}{ll}
|A\rangle|0\rangle&\quad\mbox{if the outcome is $\hat{n}_{0}=0$}\\
|B\rangle|1\rangle&\quad\mbox{if the outcome is $\hat{n}_{0}=1$}
\end{array}
\right.
\end{eqnarray*}

(II) 
Once the halt qubit is set to the state $|1\rangle$, 
the quantum Turing machine no more changes 
the halt qubit nor the tape string \cite{note:924a}, i.e., 
\begin{equation}\label{eq:325a}
U|q_{C}\rangle|h_{C}\rangle|T_{C}\rangle|1\rangle
=\sum_{q,d}c_{q,d}|q\rangle|h_{C}+d\rangle|T_{C}\rangle|1\rangle
\end{equation}
for any configuration $(q_{C},h_{C},T_{C})$
with $\hat{n}_{0}=1$, where
\begin{equation}
c_{q,d}=\delta(q_{C},T_{C}(h_{C}),T_{C}(h_{C}),q,d).
\end{equation}

(III) After the measurement of the halt flag $\hat{n}_{0}$ gives
the outcome $1$, the tape string $\hat{T}$ is measured
and the outcome of this measurement is defined to be the {\em output}
of the computation.

In order to prove that the measurement of the halt flag $\hat{n}_{0}$
does not spoil the computation, it suffices to prove that
the probability distribution of the output is not 
affected by monitoring of the halt flag.
Let $P$ be the spectral projection of $\hat{n}_{0}$
corresponding to the eigenvalue $1$, i.e.,
\begin{equation}\label{eq:328f}
P=I\otimes I\otimes I\otimes|1\rangle\langle{1}|.
\end{equation}
Since the tape is always filled with a {\em finite} sequence 
of symbols from a finite set of available symbols other than
blank cells,
the number of all the possible tape strings is countable,
so that we assume them to be indexed as $\{T_{1},T_{2},\ldots\}$.
Thus, the observable $\hat{T}$ describing the tape string can be
represented by
\begin{equation}
\hat{T}=\sum_{j=1}^{\infty}
\lambda_{j}I\otimes I\otimes|T_{j}\rangle\langle{T_{j}}|\otimes I,
\end{equation}
where $\{\lambda_{1},\lambda_{2},\ldots\}$ is a countable set of 
{\em positive}
numbers in one-to-one correspondence with $\{T_{1},T_{2},\ldots\}$
\cite{note:924b}.
Let $Q_{j}$ be the spectral projection of $\hat{T}$ pertaining to $\lambda_{j}$,
i.e.,
\begin{equation}\label{eq:328g}
Q_{j}=I\otimes I\otimes|T_{j}\rangle\langle{T_{j}}|\otimes I.
\end{equation}
We shall write $P^{\perp}=I-P$ and
$Q_{j}^{\perp}=I-Q_{j}$.

Let $\Pr\{\mbox{output}=T_{j}|\mbox{monitored}\}$ be the probability of 
finding the output $T_{j}$ up to $N$ steps by the halt scheme.
Let $\Pr\{\mbox{output}=T_{j}|\mbox{not-monitored}\}$ be the probability 
of finding the output $T_{j}$ by the single measurement after $N$ steps.  
We shall prove
\begin{eqnarray}
\lefteqn{\Pr\{\mbox{output}=T_{j}|\mbox{monitored}\}}\quad\nonumber\\
&=&
\Pr\{{\mbox{output}=T_{j}}|\mbox{not-monitored}\}\label{eq:328l}
\end{eqnarray}
Let $\psi$ be an arbitrary state vector.
If $\psi$ is the state of the machine before the computation,
we have
\begin{equation}
\Pr\{\mbox{output}=T_{j}|\mbox{not-monitored}\}
=\|PQ_{j}U^{N}\psi\|^{2}.
\end{equation}
By the projection postulate, the joint probability of obtaining the
outcome $\hat{n}_{0}=0$ at the times $1,\ldots,K-1$ and 
obtaining the outcomes $\hat{n}_{0}=1$ and $\hat{T}=\lambda_{j}$ at
the time $K$ is given by 
\begin{equation}
\|PQ_{j}(UP^{\perp})^{K}\psi\|^{2}
\end{equation}
(see \cite{Wig63} for the general formula for joint probability
distribution of the outcomes of successive measurements),
and hence we have
\begin{eqnarray}
\lefteqn{\Pr\{\mbox{output}=T_{j}|\mbox{monitored}\}}\quad\nonumber\\
&=&\|PQ_{j}\psi\|^{2}+\|PQ_{j}UP^{\perp}\psi\|^{2}+\cdots\nonumber\\
& &\mbox{ }+\|PQ_{j}(UP^{\perp})^{N}\psi\|^{2}.
\end{eqnarray}
Thus, it suffices to prove the relation
\begin{eqnarray}
\|PQ_{j}U^{N}\psi\|^{2}
&=&\|PQ_{j}\psi\|^{2}+\|PQ_{j}UP^{\perp}\psi\|^{2}+\cdots\nonumber\\
& &\mbox{ }+\|PQ_{j}(UP^{\perp})^{N}\psi\|^{2}\label{eq:325e}
\end{eqnarray}
for any $N$ and any state vector $\psi$.

We first consider the case where $N=1$, i.e.,
\begin{equation}\label{eq:325f}
\|PQ_{j}U\psi\|^{2}=\|PQ_{j}\psi\|^{2}+\|PQ_{j}UP^{\perp}\psi\|^{2}.
\end{equation}
 From (\ref{eq:325a}), the range of $PQ_{j}$ is an invariant subspace
of $U$, and hence we have 
\begin{equation}\label{eq:325b}
PQ_{j}U^{K}PQ_{j}=U^{K}PQ_{j}
\end{equation}
for any $K=1,2,\ldots$.
It follows that 
\begin{equation}\label{eq:325d}
PQ_{j}UPQ_{j}^{\perp}=\sum_{k\not=j}PQ_{j}UPQ_{k}=0.
\end{equation}
 From (\ref{eq:325b}) and (\ref{eq:325d}), we have
\begin{eqnarray}
PQ_{j}U\psi
&=&PQ_{j}UPQ_{j}\psi+PQ_{j}UPQ_{j}^{\perp}\psi+PQ_{j}UP^{\perp}\psi\nonumber\\
&=&UPQ_{j}\psi+PQ_{j}UP^{\perp}\psi.\label{eq:326a}
\end{eqnarray}
 From (\ref{eq:325b}), we have
\begin{eqnarray}
\langle UPQ_{j}\psi|PQ_{j}UP^{\perp}\psi\rangle
&=&
\langle PQ_{j}UPQ_{j}\psi|UP^{\perp}\psi\rangle\nonumber\\
&=&
\langle UPQ_{j}\psi|UP^{\perp}\psi\rangle\nonumber\\
&=&0,\label{eq:326b}
\end{eqnarray}
 From (\ref{eq:326a}) and (\ref{eq:326b}), 
we obtain (\ref{eq:325f}).

The proof for general $N$ runs as follows.
We use mathematical induction and assume that (\ref{eq:325e}) holds for $N-1$.
By replacing $\psi$ by $U^{N-1}\psi$ in (\ref{eq:325f}),
we have 
\begin{equation}
\|PQ_{j}U^{N}\psi\|^{2}
=\|PQ_{j}U^{N-1}\psi\|^{2}+\|PQ_{j}UP^{\perp}U^{N-1}\psi\|^{2}.
\label{eq:325g}
\end{equation}
 From (\ref{eq:325b}), we have
$P^{\perp}UP=\sum_{j}P^{\perp}UPQ_{j}=0$,
and hence 
$P^{\perp}U=P^{\perp}UP^{\perp}$
so that $P^{\perp}U^{N-1}=P^{\perp}(UP^{\perp})^{N-1}$.
It follows that 
\begin{equation}\label{eq:326c}
\|PQ_{j}UP^{\perp}U^{N-1}\psi\|^{2}=\|PQ_{j}(UP^{\perp})^{N}\psi\|^{2}.
\end{equation}
By induction hypothesis, we have
\begin{eqnarray}
\|PQ_{j}U^{N-1}\psi\|^{2}
&=&\|PQ_{j}\psi\|^{2}+\|PQ_{j}UP^{\perp}\psi\|^{2}+\cdots\nonumber\\
& &\mbox{ }+\|PQ_{j}(UP^{\perp})^{N-1}\psi\|^{2}.\label{eq:326d}
\end{eqnarray}
Therefore, from (\ref{eq:325g}), (\ref{eq:326c}), and (\ref{eq:326d}), 
we obtain
(\ref{eq:325e}). The proof is completed.
	
While we have discussed the problem in the Schr\"{o}dinger
picture, in what follows we shall reformulate it in the Heisenberg 
picture to probe the related physical background.
Now we introduce a new observable $\hat{O}$ defined by
\begin{equation}
\hat{O}=\sum_{j=1}^{\infty}\lambda_{j} 
I\otimes I\otimes|T_{j}\rangle\langle{T_{j}}|\otimes|1\rangle\langle{1}|.
\end{equation}
It is easy to see that the eigenvalue 0 means that the computation
has not been completed
and that the eigenvalue $\lambda_{j}$ means that the computation
has been completed and the output is given by the 
tape string $T_{j}$;
it is natural to call $\hat{O}$ the {\em output observable}
of the quantum Turing machine.
The time evolution of the output observable $\hat{O}$ is described 
by the Heisenberg operators
\begin{equation}
{\hat O}(N)=(U^{\dagger})^{N}{\hat O} U^{N},
\end{equation}
where the time $N$ is defined to be the instant just after
$N$ steps of computation.

The problem of the validity of the halt scheme discussed previously is 
now reformulated as the following problem: Can the output observable
be measured after each step without disturbing the outcomes
of the future measurements?
This is a problem of quantum nondemolition (QND) measurement,
the notion proposed previously for the gravitational wave detection
\cite{BV74,CTDSZ80}. 
According to the theory of QND measurement,
if each measurement satisfies the projection postulate,
the condition for the successful QND measurement is 
that the Heisenberg operators are mutually commutable, i.e.,
\begin{equation}\label{eq:328e}
[{\hat O}(N),{\hat O}(N')]=0
\end{equation}
for any $N,N'$; in this case, ${\hat O}$ is called a {\em 
QND observable}.
In fact, in this case the joint probability distribution
of the repeated measurement of ${\hat O}$ at each time
is the same as the joint probability distribution of the
simultaneous measurement of the observables 
${\hat O}(1),{\hat O}(2),\ldots$, in the Schr\"{o}dinger picture,
at the initial time \cite{Lud51}.

Now I will prove that the output observable ${\hat O}$ is
a QND observable.
For any $\mu=0,\lambda_{1},\lambda_{2},\ldots$ and any $N=0,1,\ldots$,
the spectral projection $E_{N}(\mu)$ of ${\hat O}(N)$ pertaining to the
eigenvalue $\mu$ is given by
\begin{eqnarray*}
E_{N}(0)
&=&(U^{\dagger})^{N}(I\otimes I\otimes I\otimes|0\rangle\langle{0}|)U^{N},\\
E_{N}(\lambda_{j})
&=&(U^{\dagger})^{N}
(I\otimes I\otimes|T_{j}\rangle\langle{T_{j}}|\otimes|1\rangle\langle{1}|)
U^{N},
\end{eqnarray*}
where $j=1,2,\ldots$.
In order to prove that ${\hat O}$ is a QND observable, it suffices to 
prove that
\begin{equation}\label{eq:328h}
[E_{N}(\lambda_{j}),E_{0}(\lambda_{k})]=0
\end{equation}
for any $N$ and any $j,k$.
 From (\ref{eq:328f}) and (\ref{eq:328g}) we have
\begin{equation}\label{eq:328n}
E_{N}(\lambda_{j})=(U^{\dagger})^{N}PQ_{j}U^{N}.\label{eq:328d}
\end{equation}
 From  (\ref{eq:325b}) with $K=N$ and (\ref{eq:328n}), we have
\begin{equation}\label{eq:328i}
E_{N}(\lambda_{j})E_{0}(\lambda_{j})=E_{0}(\lambda_{j}).
\end{equation}
Hence, for $j\ne k$ we have 
\begin{equation}\label{eq:328k}
E_{N}(\lambda_{j})E_{0}(\lambda_{k})
=E_{N}(\lambda_{j})E_{N}(\lambda_{k})E_{0}(\lambda_{k})
=0.
\end{equation}
Thus, (\ref{eq:328h}) holds for any $j,k$, and therefore
${\hat O}$ is a QND observable.

In the following,
the validity of the halt scheme will be proved
in the Heisenberg picture.
Let $N\ge N'$.
Since $E_{N}(0)=I-\sum_{j}E_{N}(\lambda_{j})$, from (\ref{eq:328i}) and
(\ref{eq:328k}) we obtain the following relations:
\begin{mathletters}\label{eq:970925a}
\begin{eqnarray}
E_{N}(\lambda_{j})E_{N'}(\lambda_{k})&=&\delta_{jk}E_{N'}(\lambda_{j}),\\
E_{N}(\lambda_{j})E_{N'}(0)&=&E_{N}(\lambda_{j})-E_{N'}(\lambda_{j}),\\
E_{N}(0)E_{N'}(\lambda_{j})&=&0,\\
E_{N}(0)E_{N'}(0)&=&E_{N}(0).
\end{eqnarray}
\end{mathletters}

 From the above relations, for any initial state $\psi$
we have
\begin{eqnarray*}
\lefteqn{
\Pr\{{\hat O}(0)=0,\ldots,{\hat O}(K-1)=0,{\hat O}(K)=\lambda_{j}\}}\quad\\
&=&
\|E_{0}(0)\cdots E_{K-1}(0)E_{K}(\lambda_{j})\psi\|^{2}\\
&=&
\|E_{K}(\lambda_{j})\psi-E_{K-1}(\lambda_{j})\psi\|^{2}.
\end{eqnarray*}
It follows that
\begin{eqnarray*}
\lefteqn{\Pr\{\mbox{output}=T_{j}|\mbox{monitored}\}}\\
&=&
\Pr\{{\hat O}(0)=\lambda_{j}\}+\\
& &\sum_{K=1}^{N}
\Pr\{{\hat O}(0)=0,\ldots,{\hat O}(K-1)=0,{\hat O}(K)=\lambda_{j}\}\\
&=&
\|E_{0}(\lambda_{j})\psi\|^{2}+
\sum_{K=1}^{N}\|E_{K}(\lambda_{j})\psi-E_{K-1}(\lambda_{j})\psi\|^{2}\\
&=&
\|E_{N}(\lambda_{j})\psi\|^{2}\\
&=&
\Pr\{{\hat O}(N)=\lambda_{i}\}\\
&=&
\Pr\{\mbox{output}=T_{j}|\mbox{not-monitored}\}.
\end{eqnarray*}
Thus, we have proved again (\ref{eq:328l}) in the Heisenberg picture.

In (III) of the formulation of the halt scheme in this paper, it is 
assumed that the measurement to obtain the output is allowed only for 
the observable $\hat{T}$ describing directly the symbol string on the 
tape, while in Deutsch's formulation \cite{Deu85,Deu89} and in later 
work no such restriction has been taken place.
However, it is an unavoidable assumption in the definition of quantum 
Turing machine.  
In fact, if this assumption is dropped, any function is computable 
without any computational time.  To see this, suppose that the tape
strings are encoded by the natural numbers.  Let $|T_{n}\rangle$ be the 
computational basis state, ignoring the inessential degeneracy, 
in which the tape string is the one encoded by $n$ and 
let $\hat{T}$ be the observable such that $\hat{T}|n\rangle
=n|T_{n}\rangle$.  Only such $\hat{T}$ is allowed to measure for 
obtaining the output.  Otherwise, given any function $f$ of the 
natural numbers and a natural number $n$, if one prepares the tape 
in the state $|T_{n}\rangle$ 
and measures the observable $f(\hat{T})$, one gets $f(n)$ surely without 
any computation.  This contradicts the Church-Turing principle.
Thus, we cannot allow even the measurement of $f(\hat{T})$ unless
$f$ is a polynomial time computable function. 
In order to maintain arguments which do not follow the present
assumption, it is worth noting that the measurement of another 
observable $\hat{A}$ to obtain the output is justified {\em with additional 
computational steps $n$}, if $\hat{A}$ can be represented as 
$\hat{A}=(U^{\dagger})^{n}\hat{T}U^{n}$.

Bernstein and Vazirani \cite{BV93} required the synchronization of 
computational paths so that every computational path reaches a final 
configuration simultaneously.  
Since the time evolution may spoil the output of computation, 
the output cannot be observed unless the exact halting time is 
previously known.  Thus, the halt scheme should be required for
the observability of the output even when the paths are synchronized.

\acknowledgments
I thank Harumichi Nishimura for several discussions on quantum
computers.


\begin{references}\itemsep=0in
\bibitem{Sho94}
P.~W. Shor,
in {\it Proceedings of the 35th Annual Symposium on Foundations of
  Computer Science}, edited by S. Goldwasser, p.~124 
(IEEE Computer Society Press, Los Alamitos, CA, 1994).

\bibitem{Sti95}
D.~R. Stinson,
{\it Cryptography: Theory and Practice},
(CRC Press, Boca Raton, FL, 1995).

\bibitem{EJ96}
A. Ekert and R. Jozsa,
Rev.\ Mod.\ Phys.\ {\bf 68}, 733 (1996).

\bibitem{Deu85}
D. Deutsch,
Proc.\ R. Soc.\ London A {\bf 400}, 97 (1985).

\bibitem{BV93}
E. Bernstein and U. Vazirani,
SIAM Journal on Computing {\bf 26}, 1411 (1997).

\bibitem{Deu89}
D. Deutsch,
Proc.\ R. Soc.\ London A {\bf 425}, 73 (1989).

\bibitem{Yao93}
A. Yao,
in {\it Proceedings of the 34th Annual Symposium on Foundations of
  Computer Science}, edited by S. Goldwasser, p.~352 
(IEEE Computer Society Press, Los Alamitos,  CA, 1993).

\bibitem{Tur36}
A.~M. Turing,
Proc.\ London Math.\ Soc. (2) {\bf 442}, 230 (1936).

\bibitem{Mye97}
J.~M. Myers,
Phys.\ Rev.\ Lett.\ {\bf 78}, 1823 (1997).

\bibitem{Jos91}
R. Jozsa,
Proc.\ R. Soc.\ London A {\bf 435}, 563 (1991).

\bibitem{note:924a}
If the relevant outcome of the computation is designed to be 
written by the program in a restricted part of the tape, this condition
can be weakened so that the quantum Turing machine may change the
part of the tape string except that part of the tape.

\bibitem{note:924b}
More precisely, the tape strings $T_{1},T_{2},\ldots$ are encoded 
as the numbers $\lambda_{1},\lambda_{2},\ldots$ by a polynomial time
computable encoding function from the finite strings of available 
symbols to positive numbers.

\bibitem{Wig63}
E.~P. Wigner,
Am.\ J. Phys.\ {\bf 31}, 6 (1963)
[in {\em Quantum Theory and Measurement}, edited by J. A. Wheeler and
  W. H. Zurek, p.~324 (Princeton UP, Princeton, NJ, 1983)].

\bibitem{BV74}
V.~B. Braginsky and Yu.~I. Vorontsov,
Uspehi Fiz.\ Nauk {\bf 114}, 41 (1974)
[Sov.\ Phys.\ Usp.\ {\bf 17}, 644 (1975)].

\bibitem{CTDSZ80}
C.~M. Caves, K.~S. Thorne, R.~W.~P. Drever, V.~D. Sandberg, and M. Zimmermann,
Rev.\ Mod.\ Phys. {\bf 52}, 341 (1980).

\bibitem{Lud51}
G. {L\"{u}ders},
Ann.\ Physik (6) {\bf 8}, 322 (1951).

\end{references}
\end{document}